\documentclass[12pt,twoside]{article}
\usepackage{graphicx}
\usepackage{citesort}
\usepackage{ulem,fancyheadings}
\usepackage[indent]{caption2}

\makeatletter
\renewcommand\@openbib@code{% Move bibitems close together
     \advance\leftmargin\bibindent
     \itemindent -\bibindent
     \listparindent \itemindent
     \parsep -0.8ex %original value: \z@
     }
\renewcommand\section{\@startsection {section}{1}{\z@}%
                                   {-4.3ex \@plus -.2ex \@minus -.2ex}%
                                   {2.1ex \@plus.2ex}%
                                   {\normalfont\underline}}
\renewcommand\subsection{\@startsection{subsection}{2}{\z@}%
                                     {-2.1ex\@plus .1ex \@minus -.1ex}%
                                     {2.1ex \@plus .2ex \@minus -.1ex}%
                                     {\normalfont\underline}}
 \makeatother
 
\begin{document}
 \pagestyle{fancy}
 \setlength{\headrulewidth}{0pt}
 \lhead{}
%add pagenumbers in the upper right corner of the paper
 \rhead{\begin{picture}(0,0)\put(120,70){\it\thepage}\end{picture}}
 \chead{}
 \cfoot{} %no page numbers in the footer
 \hphantom{.}\vskip10mm
\begin{center}
{WIDE RANGE DIELECTRIC SPECTROSCOPY\\ ON GLASS-FORMING MATERIALS:\\ AN EXPERIMENTAL OVERVIEW}
\end{center}
\vskip10.5mm
\begin{center}
 U.~SCHNEIDER, P.~LUNKENHEIMER, A.~PIMENOV, R.~BRAND, A.~LOIDL\\

 Experimentalphysik V, Universit\"{a}t Augsburg, D-86135 Augsburg, Germany
\end{center}
\vskip10.5mm

\noindent
 \underline{Abstract} Dielectric spectroscopy is one of the most commonly used techniques for the
investigation of the dynamic response of glass-forming materials. The tremendous slow-down of the particle motions when
approaching the glass transition and especially the fast processes in glass-forming materials, which have come into the
focus of scientific interest recently, make the investigation in a wide frequency range highly desirable. Recently,
results from broadband dielectric spectroscopy on glass-forming materials in their liquid and supercooled-liquid state,
covering more than 18 decades of frequency have been reported by our group. In the present paper, we give an overview
of the various experimental setups and techniques used to collect these spectra. As an example, spectra of the
prototypical glass former glycerol are presented.
 \vskip 10mm

 \vskip 14mm

\section{INTRODUCTION}
 Despite a long history of research, the theoretical explanation of the glass transition is still one of the most challenging tasks
of modern condensed matter physics. In recent years, a variety of novel approaches to the glass transition, both
theoretical and phenomenological (e.g. \cite{mctrev,NgaiKiv}), stimulated new experimental investigations especially of
the high-frequency dynamics of glass-forming liquids (e.g.,
\cite{Men,nsrev,Cum,Wutt,Lunkigly,Lunkiorl,Lunkikyockn,Sch98,Sch99,Lunky99b}). Among these, the mode coupling theory
(MCT) \cite{mctrev} is currently most controversially discussed, explaining the glass transition in terms of a dynamic
phase transition at a critical temperature $T_{c}$ significantly above the glass temperature $T_{g}$. For frequencies
in the GHz--THz region, excess contributions are predicted for the imaginary part of a generalized susceptibility,
which up to now were mainly investigated by neutron and light scattering experiments \cite{nsrev,Cum,Wutt}. Only
recently, by combining various techniques, our group was able to obtain continuous dielectric spectra on glass-forming
materials extending well into the relevant region \cite{Lunkigly,Lunkiorl,Lunkikyockn,Sch98,Sch99,Lunky99b}. For
glycerol and propylene carbonate, spectra covering 18 decades of frequency and extending well into the THz range were
obtained \cite{Sch98,Sch99,Lunky99b}. In the present paper we give a detailed description of the techniques employed to
obtain these data. Typical broadband spectra, obtained by the combination of these techniques, are shown for the
prototypical glass-former glycerol.

 \chead[U. SCHNEIDER {\it et al.}]
 {WIDE RANGE DIELECTRIC SPECTROSCOPY...}

\section{EXPERIMENTAL DETAILS}
\begin{figure}
 \centering
 \includegraphics[clip,height=10cm,width=6cm,angle=-90]{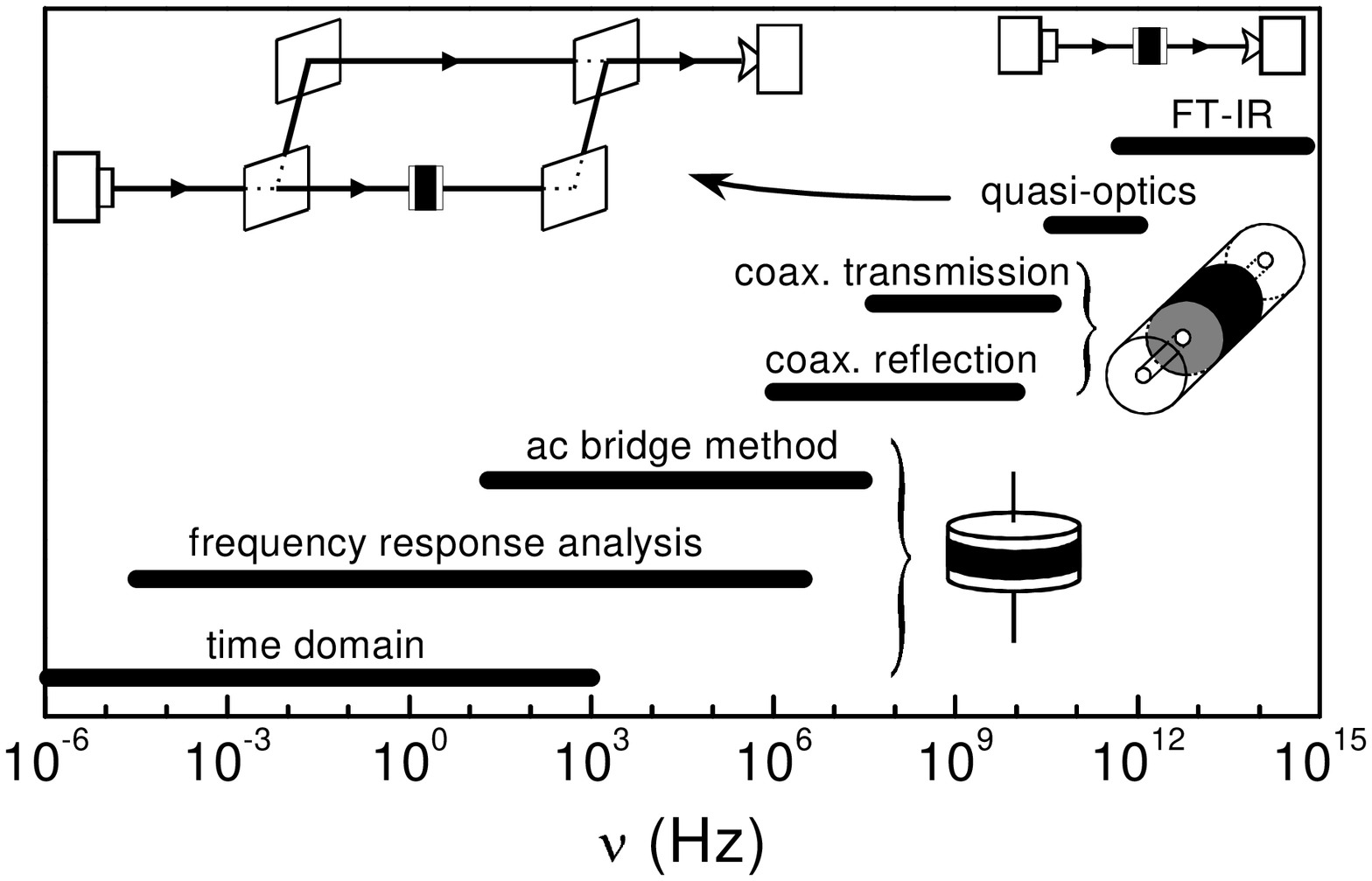}
 \caption{Overview of the employed techniques and corresponding frequency
ranges. The sample geometries and set-ups of the spectrometers are schematically indicated in the figure.}
 \label{fig:setups}
\end{figure}

Figure~\ref{fig:setups} gives an overview of the different techniques used in our laboratory. An extraordinary wide
range of frequencies, continuously covering nearly 21 decades from 10$^{-6}$ to 10$^{15}$\thinspace Hz, can be
accessed.
For cooling and heating of the samples various cryo\-stats and ovens are used. The samples have to be supercooled with
a sufficiently large cooling rate and polished sample cells have to be used to avoid crystallization. In case of an
enhanced crystallization tendency it is necessary to cool directly from a temperature above the melting point to the
measurement temperature.

To cover the complete frequency range, a single curve of real or imaginary part of the dielectric permittivity,
$\varepsilon ^{\ast }=\varepsilon ^{\prime }-i\varepsilon ^{\prime \prime}$, at a given temperature is composed using
results from different setups. For some techniques the absolute values of $\varepsilon ^{\prime}$ and $\varepsilon
^{\prime \prime}$ are not always well defined. Possible sources of error are, e.g., stray capacitances, ill-defined
sample geometries or an incomplete filling of the capacitors and sample holders. In most cases the frequency ranges of
the different devices overlap. This facilitates the matching of the curves by the application of scaling factors. In
general, only one scaling factor per measurement series (with one sample, sample holder, calibration, etc.) should be
applied. Overall, a sufficient number of measurements, overlapping in frequency and temperature, has to be collected to
minimize the errors in the composition of broadband dielectric spectra.

\subsection{Low frequency techniques}
\label{sec:lowfrequency} In the low frequency range capacitance and conductance of the sample are measured directly.
The material is filled into a stainless steel capacitor of known geometry: the design consists of two circular plates
kept at a distance with glass-fiber spacers or an outer support ring (see icon in Figure~\ref{fig:setups}). For the
measurement of low losses, large areas of the plates and small distances between them are desirable to enhance the
measurement resolution. With polished surfaces and glass fibers of a few 10\thinspace $\mu $m in diameter, geometric
capacitances of some 100\thinspace pF can be reached.

The lowest frequencies (1\thinspace $\mu $Hz $<\nu <1$\thinspace kHz) are reached with a home made time-domain
spectrometer (TDS)~\cite{Boe95}. The time-dependent charging and discharging of the sample capacitor is measured after
applying a short voltage pulse or step. Laplace transformation of the recorded response leads to frequency-dependent
spectra. In the setup a \textit{Keithley} electrometer measures a voltage that is proportional to the charge of the
sample capacitor. Alternatively an amplifier is used, the output of which is recorded directly by an AD-converter in a
computer.

In the frequency region of 30\thinspace $\mu $Hz $<\nu <3$\thinspace MHz the so-called \textit{frequency response
analysis} is applied, measuring the voltage and current through the sample using lock-in techniques. In our laboratory
an impedance analyser Solartron Schlumberger SI1260 in conjunction with a Chelsea Dielectric Interface and a
Novovontrol alpha-analyzer are available. At higher frequencies, $10$\thinspace Hz $<\nu <30$\thinspace MHz, the
standard ac-bridge method is employed using the Hewlett-Packard LCR-meters HP4284A and HP4285A.

\vspace*{3mm}
\subsection{Coaxial method}
\begin{figure}[htb]
 \centering
\includegraphics[clip,width=10cm]{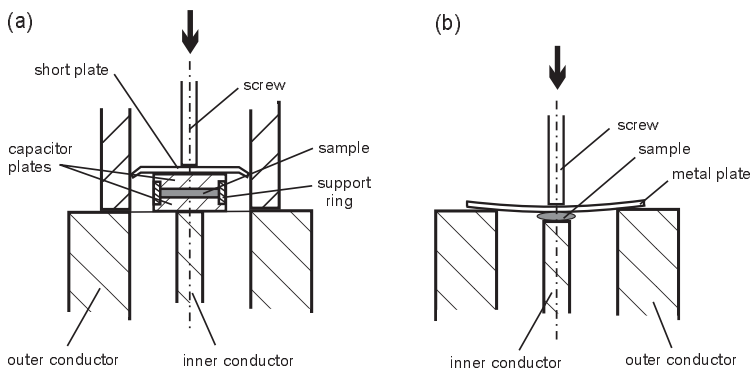}
\caption{Schematic drawing of the sample geometry for reflection measurements using the parallel plate capacitor (a)
and the droplet method (b).} \label{fig:coaxref}
\end{figure}

 In the frequency range from a few MHz to about 10\thinspace GHz a coaxial \textit{reflectometric} method is
used \cite{Boe89}. Here the sample is placed at the end of a coaxial line bridging inner and outer conductor. Either
the complex reflection coefficient is detected or a direct current-voltage measurement is performed. To correct for the
contributions of the coaxial line and connectors, a proper calibration using at least three standard impedances is
necessary. In our laboratory the \textit{Hewlett Packard} impedance analyzers HP4191A ($\nu <1$\thinspace GHz) and
HP4291A ($\nu <1.8$\thinspace GHz) and the network analyzer HP8510B ($\nu <40 $\thinspace GHz) are used. The sample can
be prepared as a parallel plate capacitor as depicted in Figure \ref{fig:coaxref}a.
Inherent to this geometry are stray capacitances and a considerable influence of the inductance of the sample holder.
These drawbacks can be avoided by applying a tiny piece or droplet of the sample to the top of the inner conductor,
which is somewhat shorter than the outer conductor (see Figure~\ref{fig:coaxref}b). By this setup losses down to
$10^{-2}$\ at frequencies up to $20$\thinspace GHz can be resolved, however with somewhat ill-defined absolute values
only.

\begin{figure}[tbp]
\centering
\includegraphics[clip,width=10cm]{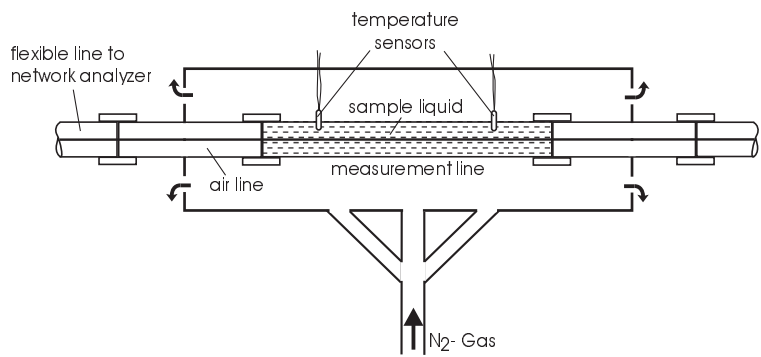}
\caption{Schematic view of the setup for temperature-dependent coaxial transmission measurements.} \label{fig:coaxtran}
\end{figure}

For higher frequencies, about $100$\thinspace MHz - $40$\thinspace GHz, the coaxial \textit{transmission} method
(Figure~\ref{fig:coaxtran}) can be applied \cite{HP85}. The sample material fills the space between inner and outer
conductor of a coaxial line which is connected to the two ports of the network analyzer HP8510B with\ flexible
extension lines. From the measured complex transmission coefficient the dielectric permittivity of the sample material
can be calculated after a proper calibration has been done. For this purpose a reflection measurement of three
standards (''open'', ''short'' and 50\thinspace $\Omega$) at the end of the extension lines and a transmission
measurement (''thru'') with both ends of the extension lines connected directly is performed. For long lines the use of
the empty sample line for the ''thru'' calibration is advantageous to correct for the signal-damping caused by line
imperfections. The transmission depends exponentially on the dielectric loss. Therefore, for different temperature
regions lines of various lengths between $10$ and $300$\thinspace mm have to be used to keep the transmission within
the resolution window of the network analyzer. To ensure a homogeneous cooling/heating of the whole line, specially
designed N$_{2}$-gas cryostats have been developed (Figure~\ref{fig:coaxtran}). Stainless-steel air lines are inserted
between the flexible and the measuring lines to ensure thermal decoupling.

\subsection{Quasi-optical spectrometer}
\begin{figure}[hbt] \centering
\includegraphics[clip,width=11cm]{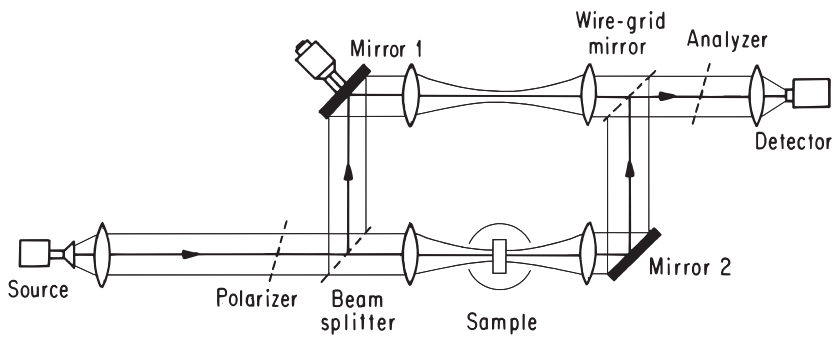}
\caption{Schematic view of the setup of the quasi-optical spectrometer; the outer thin lines denote the approximate
boundaries of the beam.} \label{fig:submm}
\end{figure}

The especially important frequency region GHz - THz (see introduction) can partly be covered by coaxial methods and
far-infrared spectroscopy, but between some 10 and some 100 GHz measurements are difficult with these techniques. In
this region resonant cavity systems can be used, but with one cavity, only a single frequency can be investigated. An
alternative is given by the free space technique where the electromagnetic wave, generated by a monochromatic source,
propagates through ''free space'' (i.e. is unguided) and is detected by a suitable detector after passing (or being
reflected by) the sample. In principle, setups as known from optical spectrometers can be applied. In our laboratory, a
spectrometer developed in the group of A.A. Volkov \cite{Vo85,Vo89} is used (Figure~\ref{fig:submm}). Its quasi-optical
setup is that of a Mach-Zehnder interferometer. It allows to measure the frequency dependence of both, the transmission
and the phase shift of a monochromatic electromagnetic beam through the sample. A frequency range $40$\thinspace GHz to
$1.2$\thinspace THz is covered continuously by a set of so-called backward-wave oscillators (BWOs) which emit a
mo\-no\-chro\-ma\-tic electromagnetic wave, tunable over a limited frequency range (typically by a factor of 2).

For the measurement the sample is put into specially designed cells made of polished stainless steel with thin
plane-parallel quartz windows. Depending on the range of frequency and temperature, the thickness of the sample cell is
chosen between $1$\thinspace mm and $30$\thinspace mm to ensure a transmission within the resolution limits of the
spectrometer.
 The beam coming from the source is parallelized with a Teflon or polyethylene lens. For the measurement
of the transmission coefficient, only the front arm of the spectrometer is used where a set of two lenses creates a
focus at the position of the sample in the cryostat or oven. The radiation is detected by a Golay cell or a pumped
He-bolometer. The noise level is reduced by the use of a chopper in the beam path and lock-in techniques for detection.
Two consecutive frequency sweeps with and without the sample are made at each temperature leading to a precise
determination of the transmission coefficient $T$. Values down to about $T=10^{-7}$ can be detected in this way. For
the determination of the phase shift, after passing a polarizer the beam is split into two with a \thinspace
$45^{\circ}$ wire-grid. In the second (reference) arm of the spectrometer a second set of lenses is situated to
compensate for the phase shift of the lenses in the from arm. After combination of the two split parts by another
$45^{\circ }$-grid and interference at the analyzer, the radiation is detected. In a reference measurement without
sample, the mirrors are adjusted for identical optical lengths of both arms by checking for the first-order minimum of
interference. The frequency-dependent phase changes are compensated automatically by a step-motor moving the mirror in
the reference arm. After repeating the measurement with the sample, the frequency dependent phase shift caused by the
sample can be calculated from the monitored mirror movements.

From transmission and phase shift the complex permittivity can be calculated by formul{\ae} for optical multilayer
interference \cite{Bo80}. Here the thickness and optical parameters of the windows have to be taken into account, which
can be determined by the measurement of the empty sample cell. The minimum value of $\varepsilon ^{\prime \prime }$
detectable with this spectrometer is only limited by the maximum thickness of the sample that can be used. In our group
values down to $\varepsilon ^{\prime \prime }=10^{-2}$ have been measured.

\vspace*{3mm}
\subsection{Fourier-transform spectrometer}
In the frequency region of $400$\thinspace GHz - $1500$\thinspace THz commercially available Fourier-transform
spectrometers (\textit{Bruker 113v} and \textit{66v/S}) are employed. Numerous sources, beam-splitters and detectors
are used to measure transmission and reflection spectra. Up to now in the investigation of glass formers we reached
frequencies up to $5$\thinspace THz, which is the upper limit when using quartz plates as sample-cell windows. If
patterns of standing waves are observed in the spectra one can unequivocally determine both the real and the imaginary
part of the complex permittivity~\cite{Bo80}. Otherwise the measured spectra are analyzed by a Kramers-Kronig
transformation to obtain the complex permittivity.

\section{MEASUREMENT RESULTS} 
\begin{figure}[bt]
\centering
\includegraphics[clip,width=10cm]{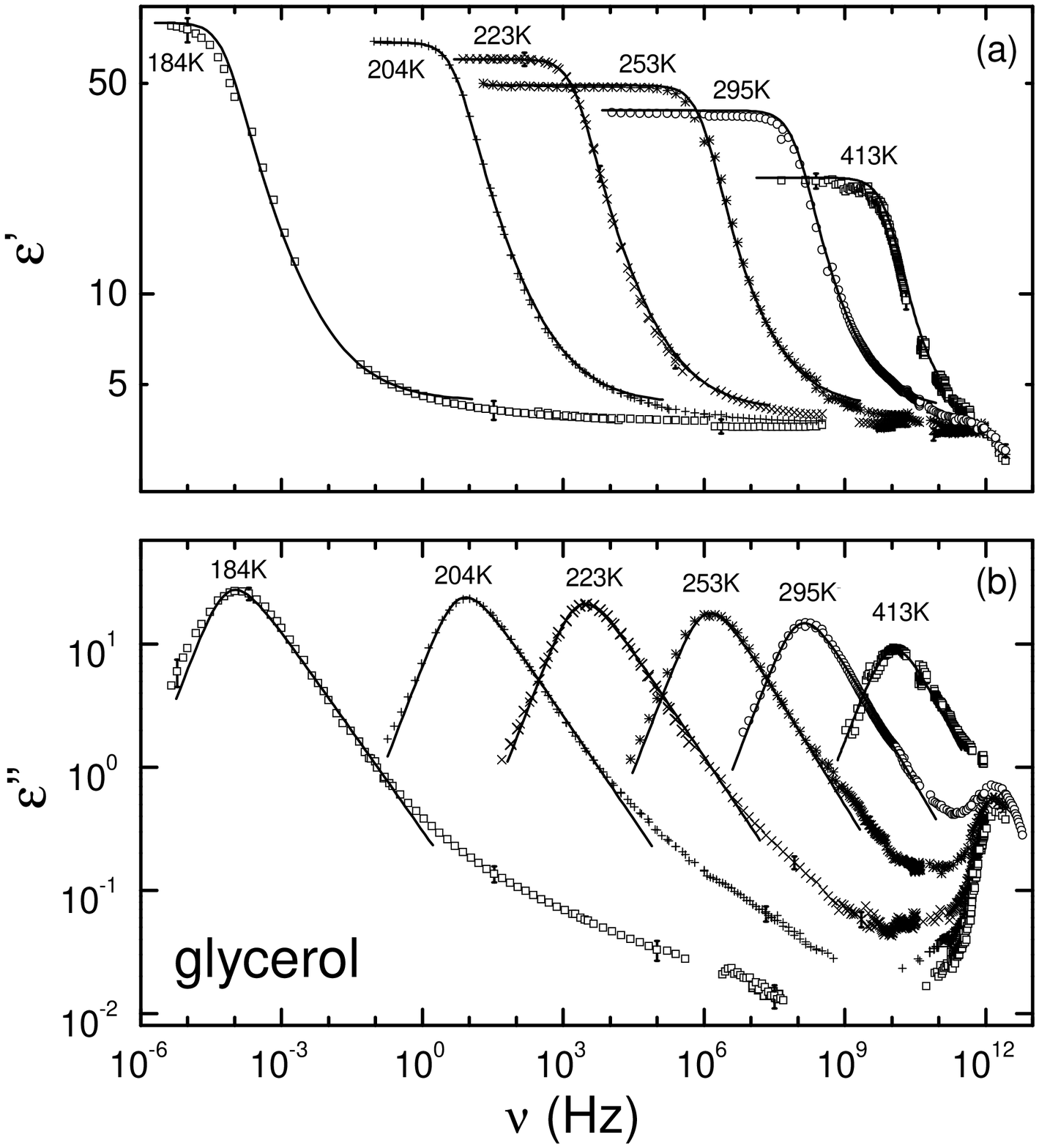}
\caption{Combined spectra of glycerol for $\protect\varepsilon^{\prime}$ (a) and $\protect\varepsilon^{\prime \prime}$
(b) for various temperatures~\protect\cite{Sch98}. The lines are fits with the Cole-Davidson
function~\protect\cite{Sch98}.} \label{fig:gle1e2}
\end{figure}

In Figure~\ref{fig:gle1e2} we present the combined results of all measurement
techniques for the prototypical glass former glycerol~\cite {Sch98}.
The curves reveal the typical behavior of glass formers: A peak in $\varepsilon^{\prime \prime}$ shows up, that
reflects the well known structural $\alpha$-relaxation which exhibits a dramatic change of time-scale with temperature.
The $\alpha$-relaxation is followed by a second power law on the high-frequency side of the $\alpha${\thinspace }-peak
(the ''excess wing''). In the GHz -\ THz region, \ not accessed in earlier dielectric investigations, a shallow
$\varepsilon^{\prime \prime}(\nu )$-minimum is observed, which is of high relevance for theoretical investigations of
the glass transition~\cite{mctrev}. Finally, near $1$\thinspace THz the boson-peak shows up, well known from neutron
and light scattering experiments \cite{nsrev,Cum,Wutt}. For a detailed theoretical analysis of the present data, the
reader is referred to \cite{Lunkigly,Lunkiorl,Lunky99b}.

\section{SUMMARY}
The large variety of dielectric techniques employed in our laboratory was described, laying special emphasis on the
non-standard high-frequency methods. By combining these techniques, the collection of nearly continuous dielectric
spectra covering about 20 decades of frequency has become possible. In this way, for the first time the temperature
evolution of the large variety of dynamic processes present in glass-forming materials can be detected by a single
experimental technique. As an example, spectra on the prototypical glass-forming liquid glycerol extending over a
frequency range of more than 18 decades, well into the THz region, were presented.

\section{ACKNOWLEDGEMENTS}
We gratefully acknowledge the help of Yu.G. Goncharov and B.P. Gorshunov in setting up the quasi-optical spectrometer.
We thank M. Dressel for assistance in some of the quasi-optical measurements and Th. Wiedenmann for technical support.
This work was supported by the DFG Grant LO264/8-1 and partly by the BMBF, contract 13N6917.

\end{document}